# Insights on the role of the covalent Ni-O bonds in LiNiO$_2$ positive electrodes: A correlative hard X-ray spectroscopy study

Jazer Jose H. Togonon[1,2,3], Jean-Noël Chotard[3,4,5], Alessandro Longo[6,7], Lorenzo Stievano[3,5,8], Laurence Croguennec[2,3,5,*], Antonella Iadecola[5,9*]

**Author's Address**

[1] Synchrotron SOLEIL, L'Orme des Merisiers, Saint-Aubin, 91192 Gif-sur-Yvette, France
[2] Univ. Bordeaux, CNRS, Bordeaux INP, ICMCB, UMR 5026, F-33600 Pessac, France
[3] ALISTORE-ERI European Research Institute, FR CNRS 3104, F-80039 Amiens Cedex 1, France
[4] Université de Picardie Jules Verne, Laboratoire de Réactivité et Chimie des Solides (LRCS), 15 rue Baudelocque, 80000, Amiens, France
[5] RS2E, Réseau Français sur le Stockage Electrochimique de l'Energie, FR CNRS 3459, Amiens F-80039 Cedex 1, France
[6] European Synchrotron Radiation Facility (ESRF), 71, Avenue des Martyrs, Grenoble F-38000, France
[7] Istituto per lo Studio dei Materiali Nanostrutturati (ISMN)-CNR, UOS Palermo, via Ugo La Malfa 153, Palermo 90146, Italy
[8] ICGM, Univ. Montpellier, CNRS, ENSCM, 34095 Montpellier, France
[9] Sorbonne Université, CNRS, Physicochimie des Électrolytes et Nanosystèmes Interfaciaux, F-75005 Paris, France

*Corresponding authors: antonella.iadecola@synchrotron-soleil.fr, laurence.croguennec@icmcb.cnrs.fr

## ABSTRACT

The interest in Ni-rich layered oxide positive electrode materials has been increasing due to its wide applicability particularly in electric vehicles as high capacity and high energy density electrode materials. However, the Ni—O bond array which builds the overall framework and plays a critical role in the charge compensation mechanism of the material requires deeper understanding. This work presents a correlative approach elucidating the role of the local highly covalent Ni—O bonds in LiNiO$_2$ (LNO) model material. Pristine and electrochemically obtained LNO positive electrodes are analyzed using *ex situ* X-ray diffraction (XRD) and extended X-ray absorption fine structure (EXAFS) to compare the average and local structural evolution upon Li$^+$ ion de-intercalation. Insights from Ni K-edge X-ray absorption near-edge structure (XANES) and non-resonant Ni Kβ X-ray emission spectroscopy (XES) spectra are combined to track the electronic environment of Ni. X-ray Raman scattering (XRS) spectra at the Ni L$_{2,3}$-edges and O K-edge provide direct bulk electronic information with regards to the interplay between Ni 3$d$ and O 2$p$ states. The overall findings imply that O plays a significant role in the charge compensation process, contributing to the substantial negative charge transfer from the O 2$p$ orbitals, because of the covalency in the Ni—O bonds inside the NiO$_2$ framework within the edge-sharing NiO$_6$ octahedra. The utilization of complementary X-ray spectroscopy techniques clarifies the intricate electronic environment of LNO, which is helpful in understanding Ni-rich positive electrode materials and offering new insights into their covalent nature.

# 1 Introduction

The design of positive electrode materials for lithium-ion batteries has been driven towards the goal of having enhanced safety and achieving high-energy density. For instance, transition metal (TM) layered oxide positive electrode materials such as NMC (LiNi$_{1-x-y}$Mn$_x$Co$_y$O$_2$) and NCA (LiNi$_{1-x-y}$Co$_x$Al$_y$O$_2$), x $\geq$ 0.8, have been suitably formulated in line with specific target applications. [1–3] In these compounds, nickel allows for high capacity and energy density, while aluminum or manganese improves thermal stability, and cobalt stabilizes the layered structure further enhancing cyclability. [4–6] As a need arises to obtain higher capacity and energy density, Ni-rich NMC and NCA compounds primarily depend on the nickel content in the material. However, increasing the energy density often entails expanding the voltage windows to fully extract Li$^+$ ions, which can lead to several detrimental issues including bulk structural and surface phase transformations, and eventually gas release. [7–9] To mitigate these issues, it is essential to fundamentally understand the electronic environment of the local Ni-O bond and its role in the charge compensation process, which is still a subject of significant debate for Ni-rich layered oxides (Figure 1).

Towards this goal, the interest in LiNiO$_2$ (LNO) has been renewed. The understanding of the atomic and electronic structure of pristine LNO, as well as the charge compensation mechanism during Li$^+$ ion extraction, remains contentious. The charge balance arguments of LNO suggest an ionic distribution of Li$^+$, Ni$^{3+}$, and O$^{2-}$ ions. However, regardless of the synthesis conditions and methods, LNO tends to contain an excess of Ni$^{2+}$ (2z), resulting to the general formula Li$_{1-z}$Ni$_{1+z}$O$_2$ where z represents the fraction of nickel occupying the lithium layer in the structure. [10–13] This implies an average oxidation state of Ni equivalent to (3+z)/(1+z), corresponding to the zNi$^{2+}$ counterparts that are created in the Ni slab. [10,14] Consequently, this affects the Li—O and especially the Ni—O distances in the average structure. Meanwhile, the local atomic environment around Ni in LNO has its own peculiarity. Earlier theoretical studies suggested that LNO could exhibit both bond disproportionation (BD) due to mixed valence states of Ni$^{2+}$ and Ni$^{3+}$ and Jahn-Teller (JT) distortion in `the Ni$^{3+}$ system. [15] More recent computational studies propose that the ground state of LNO may be a high-entropy, bond-disproportionated glass, with JT distortion dynamically reorienting on the picosecond timescale [16,17]. A static random orientation of JT distortion was also proposed based on Neutron Pair Distribution Function analysis, [18] while an extended X-ray absorption fine structure (EXAFS) study revealed a non-cooperative JT effect in the local distortion of NiO$_6$ octahedra. [19] These hypotheses entail that varying electronic configurations of Ni exist, mainly depending on the Ni—O bond framework.

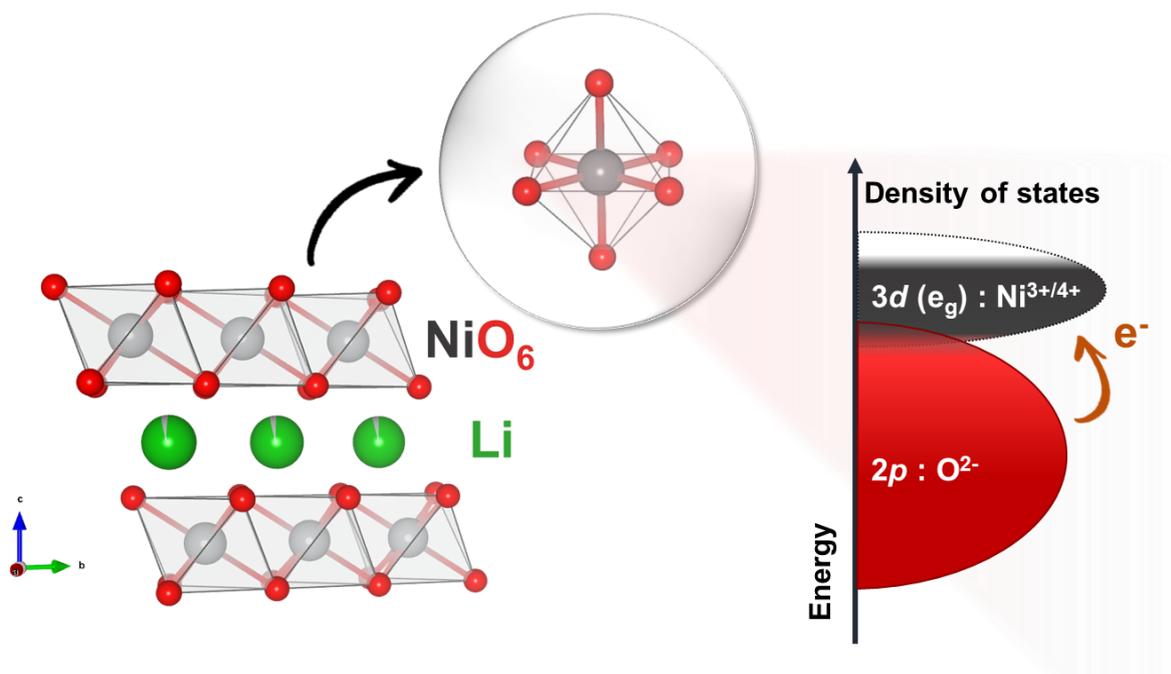

Figure 1. Representative NiO$_2$ layers made of edge-sharing NiO$_6$ octahedra and Li interlayer environment from the crystal structure of LiNiO$_2$, highlighting a local NiO$_6$ octahedra. The expected density of states of Ni and O is presented.

Since the charge compensation mechanism upon Li$^+$ ion deintercalation depends on the oxidation state of Ni in LNO, it is crucial to establish the starting point for the Ni content (i.e., the value of z) and the corresponding average Ni oxidation state in the LNO studied sample while the target here is analyzing the overall structural and electronic changes (Figure 1). During delithiation, Ni$^{3+}$ is expected to be oxidized to Ni$^{4+}$, both in their ionic low-spin configurations of $t_{2g}^6 e_g^1$ to $t_{2g}^6 e_g^0$. Moreover, it is important to note that high-voltage cycling is often attributed not only to the activation of the Ni$^{4+}$/Ni$^{3+}$ redox couple, but also to the involvement of O$^{2-}$/O$^{n-}$ anionic redox process. [20] Some studies indicate an increase in Ni—O bond covalency, with electron-hole density shifting towards oxygen when nickel is highly oxidized as there exists a strong Ni $3d$ − O $2p$ orbital overlap. [16,21,22] Others propose that dimerization of O occurs and is triggered similarly to that in Li-rich systems. [23,24]

Various spectroscopic techniques, such as Ni K-edge X-ray absorption near edge structure (XANES) [25,26] and soft X-ray absorption spectroscopy (soft-XAS) [23,27–30] at both Ni L$_{2,3}$- and O K-edges have been used to probe the charge compensation mechanism, revealing the involvement of nickel and probing the participation of oxygen. However, in the case of soft-XAS, the probing depth is limited to 5-100 nm depending on the detection mode, either electron yield or fluorescence yield, the former being mostly reported. [20,27] More recently, hard X-ray photoelectron spectroscopy (HAXPES) was used to probe the role of oxygen in the bulk charge compensation mechanism. [31] In addition to these techniques, resonant inelastic X-ray scattering (RIXS) spectroscopy was also explored to directly probe the involvement of $O^{2-}$. [20,32] Yet, limitations such as spectral overlap, low cross-section for specific oxygen-related transitions, and sensitivity to experimental conditions together with beam-induced surface artefacts [33] have constrained its ability to provide definitive insights into the role of oxygen in the redox processes. Despite these characterization tools, the lack of a comprehensive understanding of the bulk electronic structure of the material has been a major challenge in this field. Therefore, a detailed description of both the long-range and local structure, along with the electronic environment, needs to be achieved using complementary techniques on standardized samples to eliminate biases originating from the sample preparation. Moreover, the sequential evolution of such environment upon the electrochemical deintercalation of $Li^+$ ions is still underexplored.

X-ray Raman Scattering (XRS) is a suitable tool for accessing soft X-ray edges using a hard X-ray photon. In particular, it is possible to measure the O K-edge and the Ni $L_{2,3}$-edges, thus providing bulk information (> 10 µm) on charge compensation mechanisms while overcoming the limitations of soft X-ray measurements. [34,35] In fact, in the low-q geometrical configuration, the XRS spectra are consistent with those of soft-XAS, but with significantly less stringent experimental requirements. Unlike soft-XAS, XRS does not require an ultra-high vacuum sample environment, enabling *operando* measurements to be conveniently performed in a conventional electrochemical cell. [36–39] In addition, almost simultaneously measured with XRS, non-resonant X-ray emission spectroscopy (XES) at the Ni core-to-core (CtC) and valence-to-core (VtC) Kβ emission lines will can complement Ni K-edge XAS in the understanding of the Ni electronic environment, as it probes electronic transitions of occupied states close to the Fermi level, which is expected to be sensitive to valence states. [40–42]

In this study, pristine LNO and electrochemically delithiated samples at different states of charge, corresponding to various Li contents, are characterized using *ex situ* XRD and complementary hard X-ray spectroscopic techniques, including XAS, XES, and XRS. This approach provides a comprehensive and holistic view of the long-range average structure and the local structure, as well as the electronic environment of Ni and O in the materials, offering a direct probe to the nature of Ni—O bonds. The same samples are analyzed using different techniques, eliminating potential biases in the electrochemical preparation of the samples, thereby ensuring the reliability of the results. The study also aims to provide new insights into the covalent Ni—O bonds, which are directly tied to the charge compensation mechanisms in conventional stoichiometric layered oxide positive electrodes, particularly at high states of charge.

## 2 Materials and Methods

**Electrochemical cycling and sample preparation**

Tape-casted LNO positive electrodes (94:3:3 ratio of LNO, carbon black, and PVDF (BASF)) with a diameter of 14 mm were assembled in conventional 2325-coin cells in a half-cell configuration with the stack consisted of a Li metal acting as the negative and counter electrode, and Whatman glass fiber as the separator. Commercial LP30 (Solvionic, > 99.9%), consisting of 1 M $LiPF_6$ dissolved in a 1:1 v/v mixture of ethylene carbonate (EC) and dimethyl carbonate (DMC) was used as the electrolyte.

The cells were cycled at a constant current of 0.0641 mA, equivalent to a C/24 rate (where 1C corresponds to a complete charge in 1 hour), relative to the areal capacity of 1 mAh $cm^{-2}$. Charging was stopped when the target voltage, corresponding to the deintercalation of the desired amount of $Li^+$ ions was reached. The cut-off voltage and the Li content were chosen based on the expected crystallographic phase transformations demonstrated for LNO during cycling (*i.e.*, at C/24, at 3.72 V, 3.86 V, 4.04 V, 4.13 V and 4.30 V *vs.* $Li^+$/Li). When the target voltage is reached, the cells are stopped and immediately dismantled in an Ar-filled glovebox to reduce relaxation times and to preserve their state of charge and structure (complete dismantling within less than 20 mins). The recovered electrodes were washed with DMC to remove the excess of electrolyte, dried and stored in the glove box. These electrodes were then divided into four parts to have a uniform sample set for all applied characterizations. The samples are then sealed in either glass capillaries (for X-ray diffraction measurements) or Kapton films with external plastic protection (for X-ray spectroscopy measurements) to avoid air exposure during the measurements. Individual electrochemical charge profiles obtained at different cut-off voltages labelled X0 to X5 are presented in Figure S1.

**X-ray diffraction (XRD) measurements and structural analysis (Rietveld Refinement)**

*Ex situ* X-ray diffraction (XRD) patterns of pristine and cycled LNO electrodes were collected with a step size of 0.016° using a laboratory PANalytical X'Pert3 diffractometer equipped with a capillary spinner and a Cu $K\alpha_{1,2}$ X-ray source. The powders of pristine and cycled LNO electrodes were recovered by scraping off enough material from the current collector to pack a 0.30 mm diameter glass capillary.

Structural data were derived from Rietveld refinements of the XRD patterns of the pristine and cycled LNO samples using the FullProf Suite software (Supplementary Note 1). [43] Detailed crystallographic information for the pristine and its electrochemically delithiated phases is given in Tables S1.0 – S1.5 (corresponding to X0 to X5, respectively). The obtained cell parameters and the weight composition of the crystalline phases in each sample are presented in Figure S2.

**Transmission X-ray absorption spectroscopy (XAS)**

Transmission Ni K-edge (8333 eV) XAS measurements were carried out at the ROCK beamline of Synchrotron SOLEIL (Saint-Aubin, France). [44] A Ni foil was used as a reference for energy calibration. The Si(111) quick-XAS monochromator oscillated at a frequency of 2 Hz. A total of 1200 spectra were averaged over 10 minutes for each sample. The X-ray absorption near edge structure (XANES) and extended X-ray absorption fine structure (EXAFS) data collected for each sample were processed using DEMETER software package. [45]

EXAFS oscillations were $k^2$-weighted, and a sine window (dk = 0) was applied from 2.65 to 17.5 Å$^{-1}$ for the forward Fourier transform. Fitting was performed in the R range from 1 to 3 Å with a sine window (dR = 0) as well for the backward Fourier transform (Figure S3). The EXAFS parameter $S_0^2$ was set to 0.8, and $E_0$ was defined by the edge position at absorbance of 0.7 as already reported for a similar XAS study. [26]

**X-ray emission spectroscopy (XES) and X-ray Raman Scattering (XRS)**

XES and XRS measurements were carried out at ID20 beamline of the European Synchrotron Radiation Facility (Grenoble, France) with a 16-bunch filling mode. The pink beam from four U26 undulators was monochromatized to an incident energy of 9.7 keV using a cryogenically cooled Si(111) monochromator and focused to a spot size of approximately 30 μm × 30 μm (V × H) at the sample position using a mirror system in Kirkpatrick−Baez geometry. The sample was positioned to achieve a 10° grazing incident angle. All XES and XRS measurements were collected at room temperature using a custom sample holder with an aluminum dome to protect the sample from ambient air. The sample was mounted in the sample holder inside an atmosphere-controlled glovebox to avoid air exposure. Nitrogen gas flow was also introduced into the holder to further prevent air contamination. Beam attenuation was set to 10%. For XES measurements, all Kβ emission lines were collected, covering both the core-to-core (CtC) Kβ region (8110–8300 eV) and the valence-to-core (VtC) Kβ region (8300–8350 eV). Data were collected using an energy-dispersive Von Hamos spectrometer equipped with three cylindrically bent Si (440) crystal analyzers with a bending radius of R = 250 mm, and an incident energy of 9.7 keV. [46] The total integration time for each scan was approximately 15 minutes, and the data were recorded simultaneously by three different detectors. Individual scans were averaged using the PyMCA software package. [47] All spectra were background-subtracted and normalized with respect to the integrated sum of the full range of the Kβ XES. A spline-type baseline for the Kβ VtC region was created using the same approach described in ref. [48]. In short, the baseline was established using the high-energy tail of the Ni Kβ CtC emission line (8250–8300 eV) up to the high-energy end of the Kβ VtC region.

The large solid angle spectrometer at ID20 was used to collect XRS data with 36 spherically bent Si(660) crystal analyzers. Full range scans for the O K-edge were measured from 520 to 565 eV with a step size of 0.2 eV by scanning the incident beam energy to record the energy loss. The full range was divided into three regions for scanning: 520–524.8, 525–549.8 and 550–565 eV, with average acquisition times of 3.2, 37.2, and 18.75 minutes, respectively. The Ni $L_{2,3}$ edges were measured from 845 eV to 880 eV with the same step size and a total acquisition time of 43.75 minutes. Repeat scans were performed for all edges, with an average of 6–8 hours per sample, and in different zones to ensure data consistency and to avoid beam damage. All scans were checked for consistency before averaging. The data were processed using the XRStools program package as described elsewhere. [34]

# 3 Results and Discussion

## 3.1 The average crystal structure and transition metal local environment

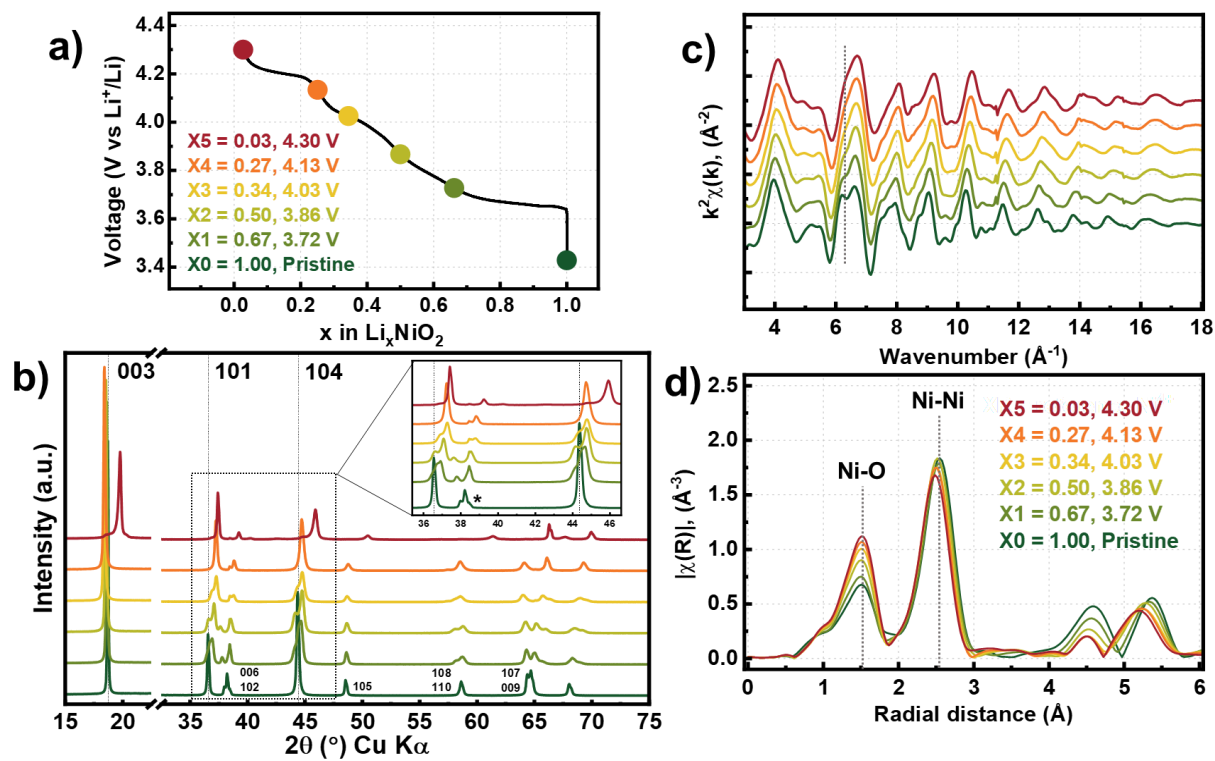

Figure 2. a) The first electrochemical charge curve obtained for the LNO electrode material *vs* Li metal, the different compositions studied in the following are highlighted and obtained at different cut-off voltages; and b) the evolution in the XRD patterns between the pristine LNO material and the electrochemically deintercalated electrodes, with inset highlighting the splitting of the 101 and 104 reflections. An asterisk in the inset is added which corresponds to the Al 111 reflections at about 38.5°; c) the $k^2$-weighted EXAFS oscillations extracted at Ni K-edge; and d) the magnitude of the Fourier transform (FT) of the EXAFS oscillations (uncorrected by the phase shift).

The first charge of LNO *vs.* Li metal at C/24 is shown in Figure 2a, highlighting different states of charge with the expected compositions. The charge profile exhibits several plateaus, characteristic of stoichiometric LiNiO$_2$ (i.e., z close to 0 in Li$_{1-z}$Ni$_{1+z}$O$_2$), which represent successive phase transitions associated with alkali and charge ordering, as well as slab gliding that leads to new stacking and structures. [2,11,25,49,50] The electrochemical curves corresponding to the cells used for the preparation of the *ex situ* samples show these specific features, demonstrating high reproducibility (Figure S1).

The XRD patterns of pristine and electrochemically deintercalated LNO electrodes are presented in Figure 2b. Rietveld refinements, performed to describe the changes in the layered structure of LNO upon Li$^+$ extraction, confirm the reliability of the materials based on reported results (see appendix for more details on the refinements, Tables S1.0 – S1.5). The refinement of the pattern of pristine LNO, sample X0, resulted in a slightly non-stoichiometric composition of Li$_{0.97}$Ni$_{1.03}$O$_2$ with 0.03 Ni$^{2+}$ in the Li site, balanced by the presence of 0.03 Ni$^{2+}$ in the Ni sites and 0.97 Ni$^{3+}$ ($z = 0.033(2)$ in Li$_{1-z}$Ni$_{1+z}$O$_2$, Table S1.0). Upon Li$^+$ deintercalation, significant changes in the XRD patterns occurred, including the emergence of a biphasic state with a new monoclinic phase (M), confirmed by the splitting of 101 and 104 reflections in Figure 2b. Further delithiation resulted in the expected phase transition back to a hexagonal phase (H), aligning with previously reported work especially at highly delithiated states. [6,13,25,50,51] The evolution of phases and their corresponding contributions are summarized in Figure S2.

Structurally, Li$^+$ deintercalation leads to the expected trends in cell parameters and cell volume. A contraction in the *a/b* cell parameters reflects the oxidation of nickel transition metal ions, inducing shortened Ni—O and Ni—Ni atomic distances (Table S2), consistent with the increasing covalency of the Ni—O framework. Meanwhile, up to the extraction of about 0.65 Li$^+$ ions, an expansion in the *c* parameter is observed despite the decreasing thickness of the NiO$_2$ slabs. This effect reflects the increased repulsive forces between oxygen layers across the interslab space as the deintercalated Li$^+$ ions no longer play their stabilization and screening role. This expansion is then followed by a decrease due to a change in the slabs stacking promoted by the covalency of the Ni—O network, as previously reported. [11,25,50]

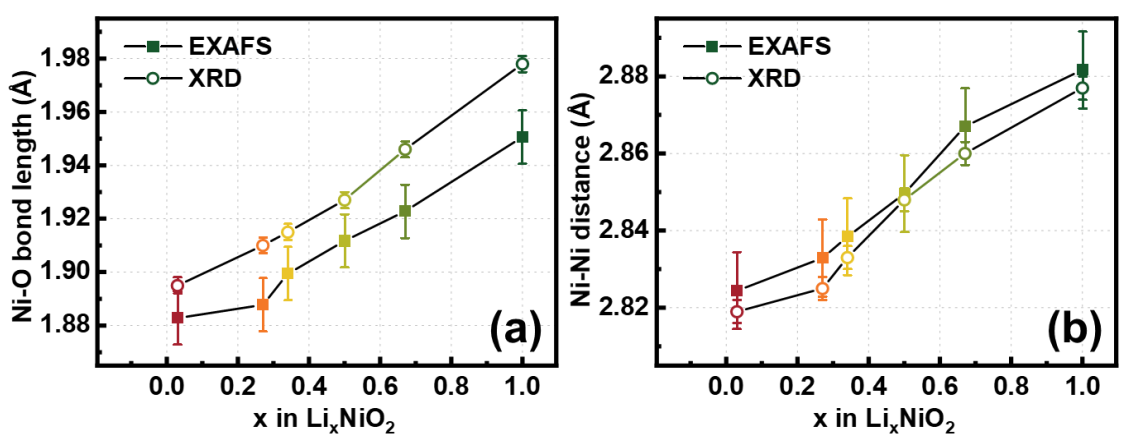

Figure 3. a) The Ni-O, and b) Ni-Ni atomic distances determined from EXAFS and XRD analyses.

The Rietveld refinements provide an average description of the Ni environment based on long-range structure, as summarized in Table S2. To this regard, it should be noted that the crystal symmetry, whether monoclinic or hexagonal, impacts the description of the local Ni structure, and in particular the average Ni—O bond length. For samples containing the two crystallographic phases, their weight fractions were used to calculate the average Ni—O and Ni—Ni distances. These values can be compared with those obtained from the analysis of the EXAFS spectra. This enables a more precise evaluation of the long-range and of the local environments of Ni, particularly in cases where local distortions might exist.

The $k^2$-weighted EXAFS oscillations show a steady evolution of the local structure as delithiation progresses (Figure 2c), especially at low wavenumbers between 5 and 10 Å$^{-1}$. These changes are reflected in the Fourier transform of the EXAFS oscillations, representing the partial local structural distribution around the absorber Ni site, where two main peaks are identified, corresponding to the Ni—O and Ni—Ni atomic distances (Figure 2d). A progressive increase in the intensity of the Ni—O peak and a slight shift to lower radial distances, accompanied by reduced Ni—Ni peak intensity, are observed. EXAFS fitting using [6] equivalent Ni—O distances for H phase-dominated samples and [4 + 2] Ni—O distances for M phase-dominated samples for the Ni—O bonds, and six equivalent Ni—Ni atomic distances lead to reasonable fits for all samples (Figure S3). This simple model should be sufficient to account for possible general structural distortions occurring in the material. [15,16,19] The distortion in the NiO$_6$ octahedra is expected for Ni$^{3+}$, characterized by a $t_{2g}^6e_g^1$ electronic configuration possibly leading to JT distortion (to a very small extent in our case, given the small z value of 0.03). Similarly, bond disproportionation (DB) due to the mixed valence between Ni$^{2+}$ ($t_{2g}^6e_g^2$) and Ni$^{3+}$ may also occur in the pristine material. The possible local distortion around Ni ([4+2] Ni—O distances) was therefore considered also in H-type structure, but its occurrence could not be clearly assessed, in line with the results of a recent XAS study. [26] Note that the Ni—O distances align well with the sum of the ionic radii (rNi$^{3+}$ = 0.56 Å, rNi$^{4+}$ = 0.48 Å, rO$^{2-}$ = 1.40 Å), especially for the pristine (~1.96 Å) and highly delithiated samples (~1.88 Å), supporting the general trend observed by both XRD and EXAFS (Figure 3).[13] A summary of the EXAFS results is shown in Table S3.

## 3.2 The charge compensation mechanism in LiNiO$_2$ and effect of covalency

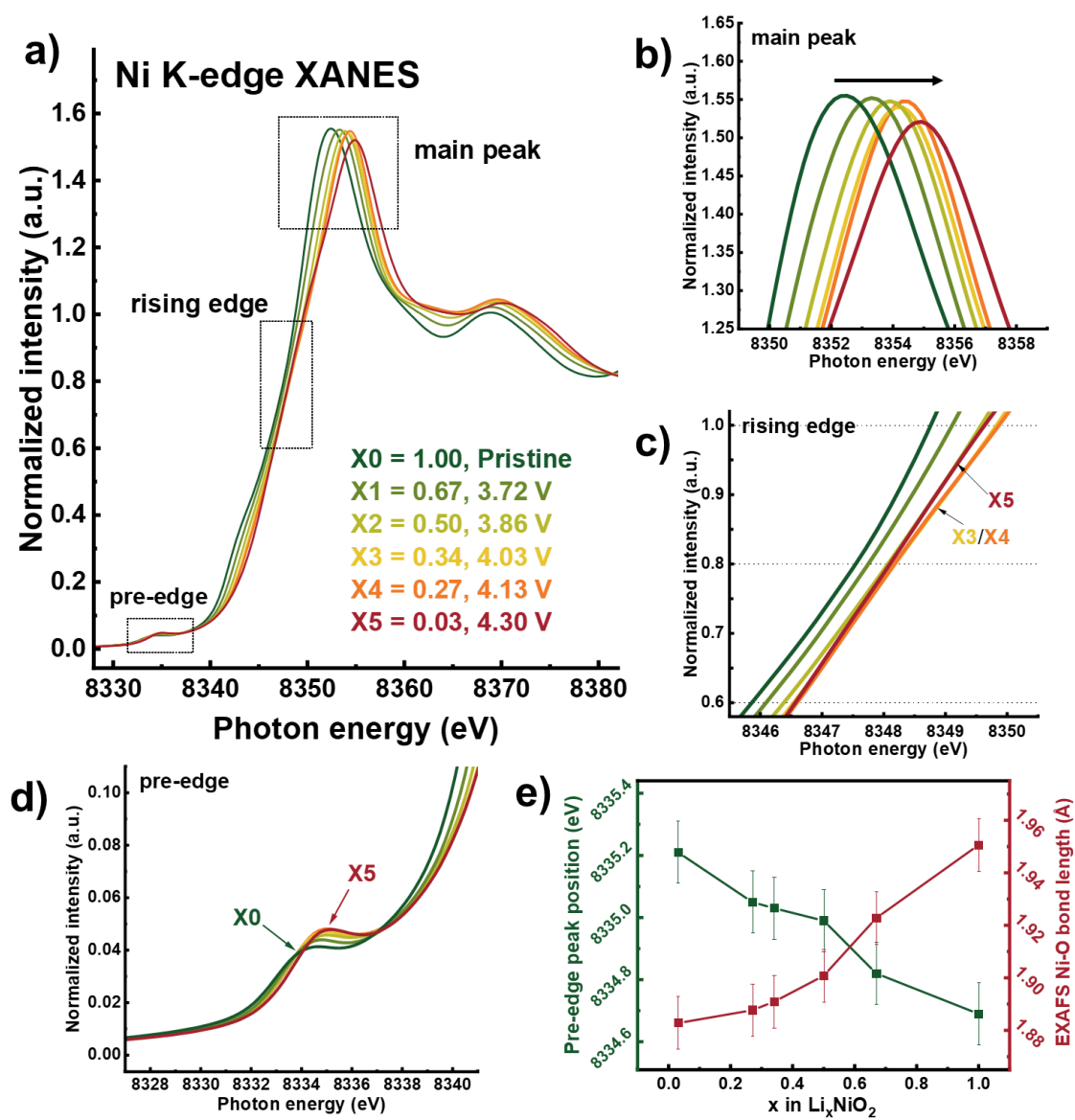

Figure 4. a) Ni K-edge XANES spectra of the pristine and the electrochemically deintercalated LNO electrodes, highlighting the evolution of b) the main peak, c) the edge position, and d) the pre-edge region; e) the correlation between the Ni—O average bond distances from EXAFS and the pre-edge peak position.

The evolution of the bulk oxidation state of Ni was followed by analyzing the *ex situ* Ni K-edge XANES spectra of the pristine and cycled LNO samples (Figure 4a). A gradual continuous shift of the main peak as well as the rising edge position towards higher energy (from 8347.5 eV to about 8348.2 eV at J = 0.8) are observed along the series, indicating the increasing oxidation of Ni upon Li$^+$ ion deintercalation. However, at cut-off voltages of 4.03 and 4.13 V (samples X3/X4), a break in the trend is observed (Figure 4b). Moreover, at the highest cutoff voltage of 4.30 V (sample X5), a different slope of the rising edge is observed (Figure 4c), indicating a more complex relationship between the rising edge position and the Ni oxidation state. Indeed, it is important to recall that the XANES features do not solely reflect the density of unoccupied states above the Fermi level, but they also depend on the local orbital rearrangement.

Meanwhile, the pre-edge intensity slightly increases and shifts to higher energy upon Li$^+$ ion extraction (Figure 4d). An increasing pre-edge intensity is expected due to the removal of electrons from nickel 3d orbitals. In addition, the increased hybridization may be directly correlated to the increasing covalent framework consequent to the shortening of the Ni—O bond distances, resulting in a mild overlap between orbitals especially when local distortions are expected (e.g., for the M phase more than for H phase). Subsequently, the observed pre-edge features support the increase in Ni oxidation state because of these local structural and electronic changes. A strong correlation is evident in Figure 4e between the pre-edge peak position and the Ni—O bond distances from EXAFS analysis as a function of the expected Li$^+$ ion composition. [6,25,50,52]

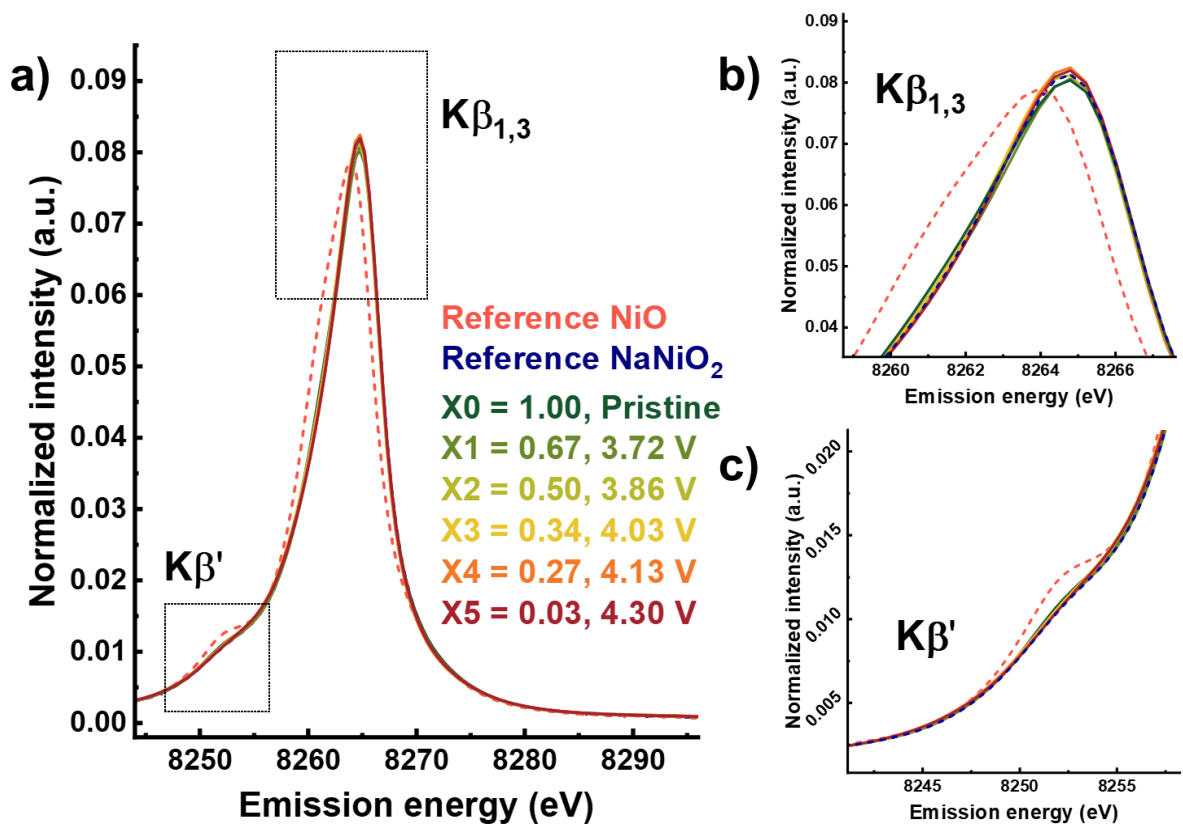

Figure 5. a) Ni Kβ X-ray emission lines of the pristine and electrochemically delithiated LNO positive electrodes, highlighting b) the CtC Kβ$_{1,3}$ region and c) the CtC Kβ' region. Reference materials, NiO (Ni$^{2+}$: $d^8$) and NaNiO$_2$ (Ni$^{3+}$: $d^7$), are included for better comparison.

While the Ni K-edge XANES energy position in the series of samples obtained upon Li$^+$ ion deintercalation generally evolves continuously with the expected oxidation of Ni, a more complex evolution is observed depending on the feature being examined, which may be likely connected to the increased covalency of Ni―O bonds. Thus, the rising edge position alone may not be sufficient to accurately determine the oxidation state of Ni, as it can be influenced by the nature of the Ni―O bonds. [23] To address this, Ni CtC and VtC Kβ XES was used as a complementary technique (Figure 5a). In fact, the Ni CtC Kβ XES line (8225 eV – 8300 eV) results from the specific electronic relaxation process corresponding to $3p \rightarrow 1s$ transition filling the Ni core-hole. The main emission line, corresponding to the CtC Kβ$_{1,3}$ region, is highly sensitive to the electronic environment, particularly the valence electrons, due to the strong $3p$―$3d$ exchange interaction, and is expected to provide insights on the electrons from the occupied states. It is apparent that a significant shift towards higher energies is observed on going from divalent NiO (~8264.3 eV) to trivalent LiNiO$_2$ (~8264.9 eV) and NaNiO$_2$ (~8264.8 eV), despite a shift towards lower emission energies is expected as oxidation state increases as similarly observed `for several transition metals (Ti, V, Cr, Mn, Fe, Co). [53,54] Limited reports have been published in the case of Ni and such trends are still under investigation. Moreover, no clear shift can be detected for the delithiated materials (Figure 5b). Additionally, it should be noted that CtC Kβ' region is also of interest due to the expected change of the overall spin state from S = ½ (Ni$^{3+}$: $d^7$) to S = 0 (Ni$^{4+}$: $d^6$). However, no visible alteration is observed suggesting that no change in the local magnetic moment across the materials relative to that of a high spin NiO with S = 1 (Ni$^{2+}$: $d^8$) (Figure 5c).

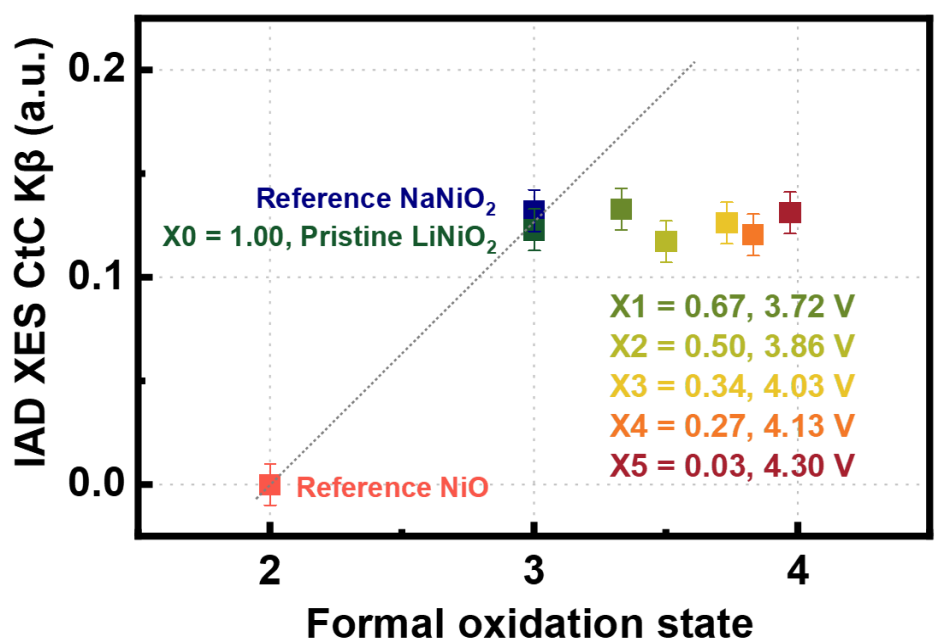

Figure 6. The result of the IAD analysis over the energy range 8200 eV – 8300 eV with the reference NiO set as IAD = 0. A line connecting the reference NiO and NaNiO$_2$ is presented as a guide for the expected slope in the increase of the formal oxidation state.

Quantitatively, the integrated absolute difference (IAD) within a range of energy can be of help for such evaluation. [42] The IAD values over a range covering the Ni CtC Kβ emission lines were calculated with the reference NiO set as the reference (IAD = 0). IAD provides insights into the local magnetic moment deriving from the electronic environment, and independent from structural effects in the material. [42] As the IAD values can directly be correlated to the unpaired electrons in the filled 3$d$ states it should also be sensitive to the overall oxidation state of Ni. [42] For simplicity, the formal oxidation state of pristine and electrochemically delithiated LNO samples, which were mainly based on ionic charge balance given that a certain amount of Li$^+$ ions was extracted, are correlated with the calculated IAD as seen in Figure 6. Note that the formal oxidation state is not an appropriate descriptor of the electronic configuration in a covalent system, as it is defined in a pure ionic picture. Interestingly, the calculated IAD values are very similar for all the Li$_x$NiO$_2$ samples (0.11 to 0.13 ± 0.01) which may suggest that the overall electronic environment in LNO does not have a significant change despite the extraction of Li$^+$ ions, confirming the covalent nature of the Ni—O bonds. [55]

On the other hand, the Ni VtC Kβ XES (8300 eV – 8350 eV) (Figure S4a) is highly sensitive to the chemical properties of the ligands coordinated to the absorber atom. [41] This emission line displays high correlation and agreement with the local bond distances obtained from EXAFS analysis (Figure S4b). [48,54,56,57] As Li$^+$ ions are extracted from the bulk LNO, a subtle but significant shift in the main peak at ~8327 eV towards higher energy is observed. The shortest Ni—O bond distance obtained from XRD analysis of ~1.8876(3) Å and from EXAFS fitting of ~1.88(1) Å corresponds to the higher emission energy (Table S2 and Table S3). The main peak of Ni Kβ$_{2,5}$ emission line is directly correlated to the average Ni—O bond distances from XRD and EXAFS refinements, probing that the VtC XES can be also used to assess on the structural evolution of positive electrode materials. [48]

Generally, the determination of the oxidation state by probing the unoccupied electronic states through Ni K-edge XANES spectra suggests an increase in Ni oxidation state, at least up to 4.03 V. Furthermore, the analysis of the occupied electronic states from Ni CtC Kβ XES spectra by IAD method indicates that the number of unpaired electrons may not actually be decreasing in the series of samples. This suggests the presence of a strong covalent interaction within the NiO$_6$ framework throughout the delithiation process. Therefore, despite having a clear decrease in bond distances between the Ni and O atoms (Table S2 and Table S3), combined results from XANES and XES suggest a strong charge transfer process between the Ni 3$d$ and O 2$p$ states, affecting the overall electronic environment and the charge compensation process upon Li$^+$ ion extraction. [21]

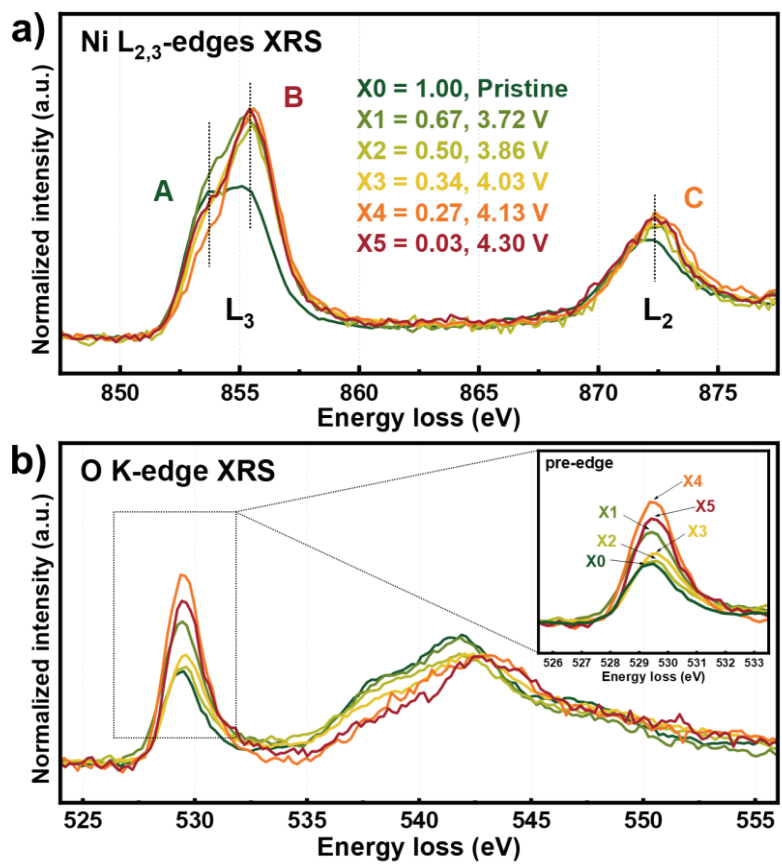

Figure 7. a) Ni $L_{2,3}$-edges and b) O K-edge XRS spectra of the pristine and electrochemically delithiated LNO positive electrodes.

The question now centers on understanding the relationship between Ni and O in LNO, considering both their oxidation states as $Li^+$ ions are extracted. In this line, it is essential to extract direct information on the Ni 3$d$-O 2$p$ electronic interaction. For this purpose, Ni $L_{2,3}$-edges XRS and the O K-edge XRS spectra were collected on the same series of electrochemically delithiated LNO electrodes (Figure 7a and Figure 7b).

The oxidation state of Ni can be assessed by looking into the Ni $L_{2,3}$-edges XRS, corresponding to the dipole electronic transition from the occupied Ni 2$p$ to the unoccupied Ni 3$d$ states. Overall characteristic features including a double peak at lower energies, at ~853.7 eV (Peak A) and at ~855.4 eV (Peak B), and another feature at higher energies at ~872.5 eV (Peak C) are consistent with Ni $L_{2,3}$-edges spectra obtained from soft-XAS in previous studies. [27,28] Specifically, the former is assigned to the Ni $L_3$-edge (Peak A and B: $2p_{3/2} \rightarrow 3d$), while the latter is for the $L_2$-edge (Peak C: $2p_{1/2} \rightarrow 3d$). Generally, a bimodal splitting feature of about 2 eV can only be observed in the $L_3$-edge and only as a slightly broad feature at the $L_2$-edge, especially for X0. Moreover, it should be noted that differences in the probing depth entails distinct features as soft-XAS is typically limited to 100 nm (using fluorescence yield detection) whereas XRS enables probing depths up to 10 µm, thus probing the bulk of the material.

[58] Upon delithiation, Peak B gradually increases in intensity, with diminishing feature at Peak A, although this trend is not linearly dependent on the amount of deintercalated Li$^+$ ions. Thus, it can be inferred that Peak B becomes more intense for more oxidized samples. This sharp main feature at the energy equivalent to ~855.4 eV is expected for Ni$^{4+}$ species, as previously reported by Jacquet and co-workers. [26] Intriguingly, a broader shoulder at Peak A is observed from sample X5, while a less pronounced feature is observed in sample X4, despite both having intense features at Peak B, which could signify partial reduction of Ni. Overall, it is compelling that the general observation in the changes in the Ni L$_{2,3}$-edges XRS would require further quantitative analysis and the possibility for theoretical studies to further elucidate the dynamic changes in the electronic environment. Nevertheless, the results obtained in probing the Ni electronic environment from the Ni K-edge XAS, Ni CtC Kβ XES, and Ni L$_{2,3}$-edge XRS confirm the complex interplay between Ni—O bonds.

The O K-edge XRS spectra exhibit two distinct regions: the pre-edge, with the most intense feature around ~529.5 eV, and a broader feature 5–10 eV above the pre-edge, corresponding to the oxygen *p*-character hybridized with the 4*s* and 4*p* orbital of Ni (Figure 7b). [59,60] The pre-edge reveals changes in the electronic environment of oxygen due to strong hybridization between the Ni 3*d* and O 2*p* orbitals. Our results reveal clear changes in the pre-edge feature upon delithiation, as seen in the increased intensity in comparison to that of the pristine (inset Figure 7b), and show more resolved features in contrast to the previously reported O K-edge soft-XAS results showing only relatively small changes in intensity. [27] In relation to the pre-edge intensity, a recent O K-edge XRS study, supported by *ab initio* calculations, reported that reproducing the intense pre-edge feature requires increased overlap between the O 2*p* and Ni 3*d* orbitals, implying higher Ni—O bond covalency. [26] The increased hybridization is clearly observed across the series, enabling electron transfer from the O 2*p* ligands to the Ni 3*d* orbitals, leaving ligand holes for lattice oxygen activation. [21,61–63] This is further supported by the charge redistribution between oxygen and nickel atoms at high voltage in the highly delithiated state, suggesting the participation of oxygen holes in the charge compensation mechanism. [26] Similar strong charge transfer effects within the hybridized Co 3*d* and O 2*p* orbitals of the Co—O bonds in LiCoO$_2$ positive electrodes, especially at high state of charge, was recently reported by Asakura and co-workers wherein a partially reduced Co environment was observed as a consequence of the presence of ligand holes in the oxygen orbitals. [64] This is in line with the possibly covalent CoO$_2$ network resulting in a negative charge transfer from O 2*p* to Co 3*d* orbitals, which plays an important role in the charge compensation mechanism.

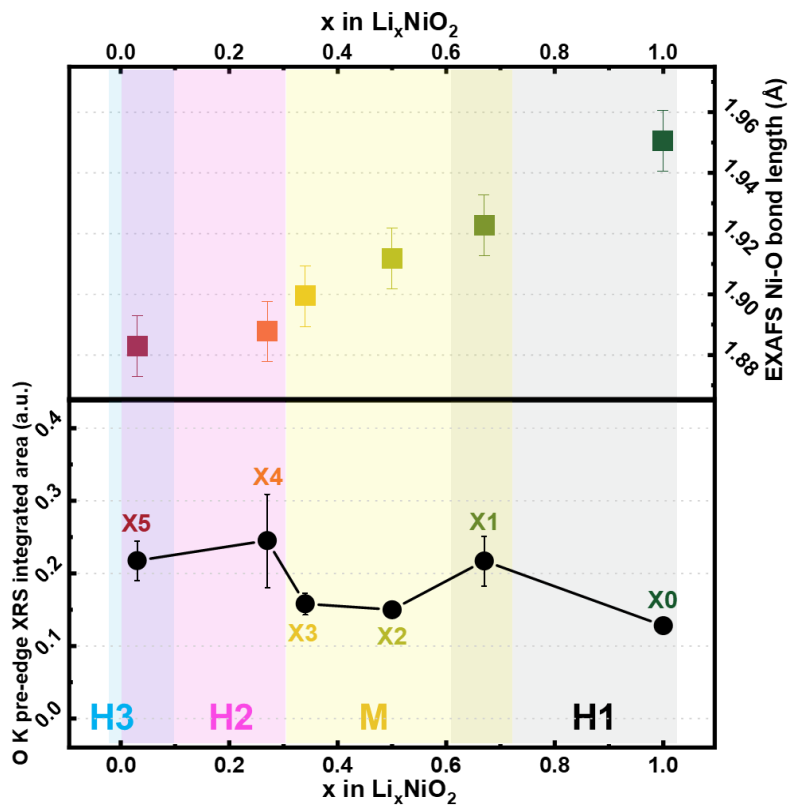

Figure 8. The correlation between (top) the local Ni-O bond distances from EXAFS analysis and (bottom) the integrated area of the pre-edge of the O K-edge XRS spectra. Highlighted areas correspond to the global crystal phases indicated by H1, M, H2, and H3.

Furthermore, the integrated oxygen XRS pre-edge area increases during $Li^+$ ion deintercalation, followed by a decrease, and then a subsequent increase (Figure 8). Notably, samples X2 and X3, both purely monoclinic with a distortion of the triangular oxygen lattice compared to the hexagonal phases, as well as X5, characterized by a different oxygen stacking (O1-type in H3 versus O3-type in H1/H2/M) [25], show a decrease in pre-edge intensity. This implies that, as phase transitions occur upon delithiation, the overall crystallographic structure may also impact, on average, the Ni—O bond covalency. Otherwise, this decrease in covalency may also be related to the local Li vacancy ordering affecting the Ni—O covalency in the local structure. [12,25] A systematic and detailed theoretical study is needed to further elucidate such a decrease in covalency in the monoclinic phase, to determine whether it is related to the long-range structure or local geometric environment, or to both effects.

These findings provide original insights into the possible relationships among structural phase transitions, electronic properties, covalency and Ni—O bond distances, shedding new light on the role of nickel and oxygen in the charge compensation process. The results suggest that changes in covalency may occur independently of the Ni—O bond distances and could mainly rely on the electronic structure of LNO. Furthermore, an induced negative charge transfer to the Ni $3d$ environment from the O $2p$ states is confirmed, producing ligand holes in the oxygen orbitals. [21,23,31]

Based on the presence of such ligand holes in LiNiO$_2$, its initial electronic configuration could be better described by 3$d^{7+n}$L$^n$ (L indicates one hole in the ligand site) states instead of the conventional low spin 3$d^7$ ones, implying the dynamic interplay between Ni and O, which influences the overall electronic environment. [23] Furthermore, as more Li$^+$ ions are deintercalated, the 3$d$ orbital of Ni lowers in energy, resulting in an increased overlap with the 2$p$ orbital of the oxide ion, further stabilizing the 3$d^{6+n}$ L$^n$ (where $0 > n > 1$) configuration and resulting in the formation of ligand holes. [21,23,26,64,65] Covalency is expected to increase with greater overlap between the 2$p$ and 3$d$ orbitals, which is typically associated with shorter metal-oxide bond lengths. This work, however, reveals a more complex interplay, showing that covalency is also influenced by the evolution of the average long-range structure.

# 4 Conclusion

A comprehensive correlative analysis of long-range and local structure of pristine and electrochemically delithiated LNO positive electrodes was carried out through *ex situ* XRD and EXAFS studies. Additional insights into the bulk electronic structure of LNO at different states of charge is obtained using complementary hard X-ray spectroscopic techniques. Moreover, the proposed experimental workflow relying on the use of the same samples and varying conditions of spectral collection excluding beam-induced damage, not only eliminates potential discrepancies in the uniformity of the materials used but also demonstrates the reliability and reproducibility of the results.

Specifically, Ni K-edge XAS allows following the evolution of the electronic environment of Ni, aligning with structural changes observed in the EXAFS analysis and consistent with XRD refinements. This highlights how the local coordination of Ni within the NiO slabs evolves during delithiation. Moreover, the findings from the Ni CtC Kβ XES through IAD analysis suggest that a strong influence on the covalency in the Ni—O bonds may exist across samples. In addition, Ni L$_{2,3}$-edge XRS further clarifies the role of Ni in the charge compensation mechanism. The strong involvement of oxygen in the charge compensation process during the delithiation is also confirmed by O K-edge XRS. The combination of these techniques allows probing the covalent nature of the Ni—O bonds, which induces a stronger reductive effect, especially at high potential. In particular, the covalent nature of the Ni—O bonds, in pristine as well as in delithiated LNO, appears very complex, being influenced not only by the local Ni—O environment but also by the long-range structure. The latter involve differences in oxygen packing, NiO$_2$ slab interlayer interaction and Li/vacancy ordering, but also possible distortions of the oxygen triangular lattice.

These findings imply that understanding the interplay between the local and the long-range structural changes is crucial for unraveling the charge compensation mechanisms in LNO. Theoretical studies, particularly focused on the relationship between the short- and the long-range structural ordering, are therefore necessary to fully describe and correlate the structural and electronic environment in similar class of materials. These studies would extend our understanding of the interaction between covalency and structural phase transitions during the delithiation process, offering valuable insights into Ni-rich positive electrode materials. This improved understanding is mandatory to shed light on their contribution to the development of more efficient and stable electrode materials.

## Acknowledgements

As part of the DESTINY PhD Programme, J.J.H.T., J.-N.C., A.I., and L.C, acknowledge the funding from the European Union's Horizon 2020 research and innovation programme under the Marie Skłodowska-Curie Actions COFUND (Grant Agreement #945357). J.J.H.T also acknowledges the funding and financial support from the Synchrotron SOLEIL. A.I., L.C., J.-N. C. and L.S. recognize the funding by the French National Research Agency (STORE- EX Labex Project ANR- 10-LABX-76-01). The authors also acknowledge the European Synchrotron Radiation Facility (ESRF) for providing beamtime at the ID20 beamline (Proposal No. CH6843). The authors are also grateful to Christoph Salhe, Blanka Detlefs and Florent Gerbon for the technical support during the experiments (ESRF, ID20). Synchrotron SOLEIL is acknowledged for providing beamtime at the beamline ROCK which also benefits from ANR grant as part of the"Investissements d'Avenir" program ANR-10-EQPX-45.

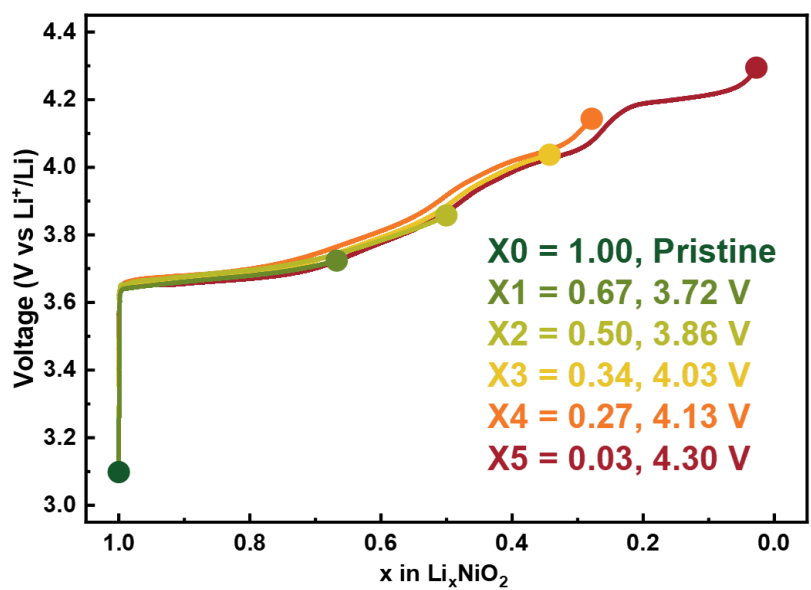

Figure S1. Electrochemical charge curves obtained for the LiNiO$_2$ positive electrode, recovered at different target voltages and representing different states of charge.

Supplementary Note 1. XRD Rietveld refinement procedures

All diffractograms obtained from XRD measurements are refined by Rietveld method. An additional phase was included to account for the Al current collector present in the powder from the tape-casted electrodes. A feature corresponding to the Al 111 at about 38.5° was detected and was refined to enhance the quality of the line profile description of the phase of interest and the fit reliability using for the Al the $Fm\bar{3}m$ cubic space group with lattice parameter 4.048 Å. Reported values in main phase contribution are normalized excluding the contribution of Al.

Meanwhile, iterative refinement procedures were done by optimizing the cell parameters, shape parameters, atomic positions, Biso and occupancy values. The Li occupancy in the pristine and *ex situ* LNO samples were set differently. Initially for the pristine, the presence of Ni in the Li layer was considered, therefore it was refined together with the amount of Li. Subsequently, this amount of Ni in the Li site was kept constant. For the *ex situ* samples, the Li content set for the refinements was based from the expected compositions obtained from the electrochemical data. This was set as a constraint due to the limitation of XRD in accounting for Li content due its low scattering.

Moreover, Biso values were also refined for the Li, Ni1, Ni2 and O1 sites for the pristine material, those of Li and Ni1 atoms being constrained to be equal since they are localized on the same site. Subsequent refinements of the Biso values for the delithiated samples were then performed, except when limited by the information that could be obtained from these data collected for *ex situ* (sometimes biphasic) samples. In that case, these Biso were fixed.

Table S1.0 Structural information including the lattice and atomic parameters of pristine LiNiO$_2$ (X0) obtained from Rietveld refinement of the powder XRD data.

| X0 – LiNiO$_2$ | |
| --- | --- |
| Phase | Hexagonal – H1 |
| Space group | R -3m |
| a = | 2.8772(3) Å |
| c = | 14.195(3) Å |
| α = β, γ | 90°, 120° |
| V = | 101.77(3) Å$^3$ |
| V/Z = | 33.92(3) Å$^3$ |

| Atom | Wyckoff | x/a | y/b | z/c | Occ. | B$_{iso}$ (Å$^2$) |
| --- | --- | --- | --- | --- | --- | --- |
| Li | 3a | 0 | 0 | 0 | 0.967(2) | 0.78(2) |
| Ni1 | 3a | 0 | 0 | 0 | 0.033(2) | 0.78(2) |
| Ni2 | 3b | 0 | 0 | ½ | 1 | 0.15(10) |
| O1 | 6c | 0 | 0 | 0.2418(2) | 1 | 0.84(6) |

| Distance | |
| --- | --- |
| Ni2–O | 1.9771(5) × 6 |
| Ni2–Ni2 | 2.8772(3) × 6 |
| Li(Ni1)-O | 2.1060(5) × 6 |

Agreement factors: R$_p$ = 3.15, R$_{wp}$ = 4.54, χ$^2$ = 4.89

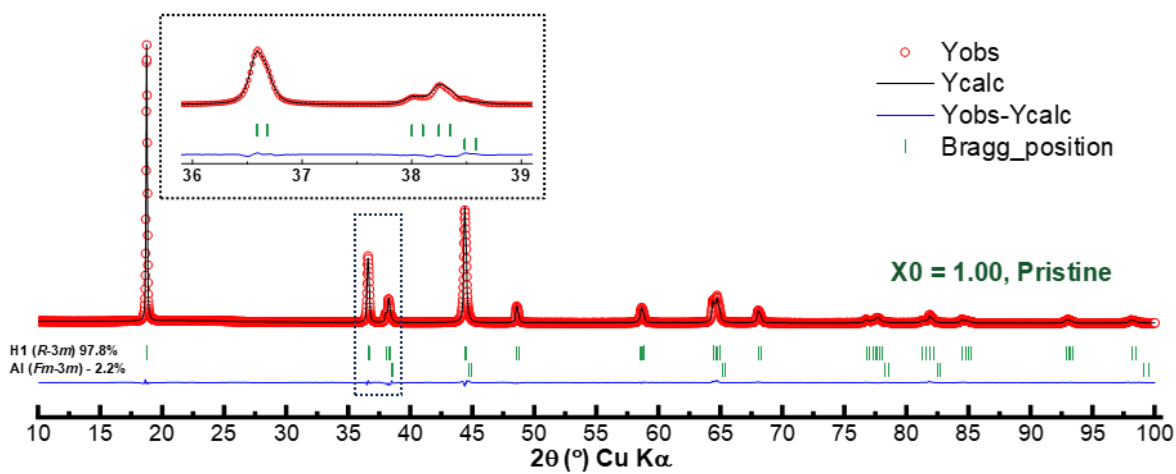

Table S1.1 Structural information including the lattice and atomic parameters of *ex situ* $Li_{0.67}NiO_2$ (X1) obtained from Rietveld refinement of the powder XRD data.

| | X1 – $Li_{0.67}NiO_2$ | | | | | |
|---|---|---|---|---|---|---|
| Phase | Monoclinic – M, 79.9 wt.% | | | | | |
| Space group | C 2/m | | | | | |
| a = | 4.9864(6) Å | | | | | |
| b = | 2.8397(3) Å | | | | | |
| c = | 5.0490(6) Å | | | | | |
| β | 109.68(8)° | | | | | |
| V = | 67.32(4) Å³ | | | | | |
| V/Z = | 33.66(2) Å³ | | | | | |
| Atom | Wyckoff | x/a | y/b | z/c | Occ. | $B_{iso}$ (Å²) |
| Li | 2d | 0 | ½ | ½ | 0.670 | 0.95 |
| Ni1 | 2d | 0 | ½ | ½ | 0.033 | 0.95 |
| Ni2 | 2a | 0 | 0 | 0 | 1 | 0.53 |
| O1 | 4i | 0.7405(5) | 0 | 0.2180(8) | 1 | 0.86 |
| | Distance | | | | | |
| Ni2–O | | | | 1.9610(9) × 2, 1.9435(2) × 4 | | |
| Ni2–Ni2 | | | | 2.8397(3) × 2, 2.8691(6) × 4 | | |
| Li(Ni1)-O | | | | 2.1480(9) × 2, 2.1156(6) × 4 | | |
| Phase | Hexagonal – H1, 20.1 wt.% | | | | | |
| Space group | R -3m | | | | | |
| a = | 2.8636(2) Å | | | | | |
| c = | 14.260(2) Å | | | | | |
| V = | 101.77(3) Å³ | | | | | |
| V/Z = | 33.65(2) Å³ | | | | | |
| Atom | Wyckoff | x/a | y/b | z/c | Occ. | $B_{iso}$ (Å²) |
| Li | 3a | 0 | 0 | 0 | 0.670 | 0.78 |
| Ni1 | 3a | 0 | 0 | 0 | 0.033 | 0.78 |
| Ni2 | 3b | 0 | 0 | ½ | 1 | 0.15 |
| O1 | 6c | 0 | 0 | 0.2390(9) | 1 | 0.84 |
| | Distance | | | | | |
| Ni2–O | | | | 1.9480(6) × 6 | | |
| Ni2–Ni2 | | | | 2.8636(2) × 6 | | |
| Li(Ni1)-O | | | | 2.1320(9) × 6 | | |

Agreement factors: $R_p$ = 5.29, $R_{wp}$ = 6.73, $\chi^2$ = 5.61

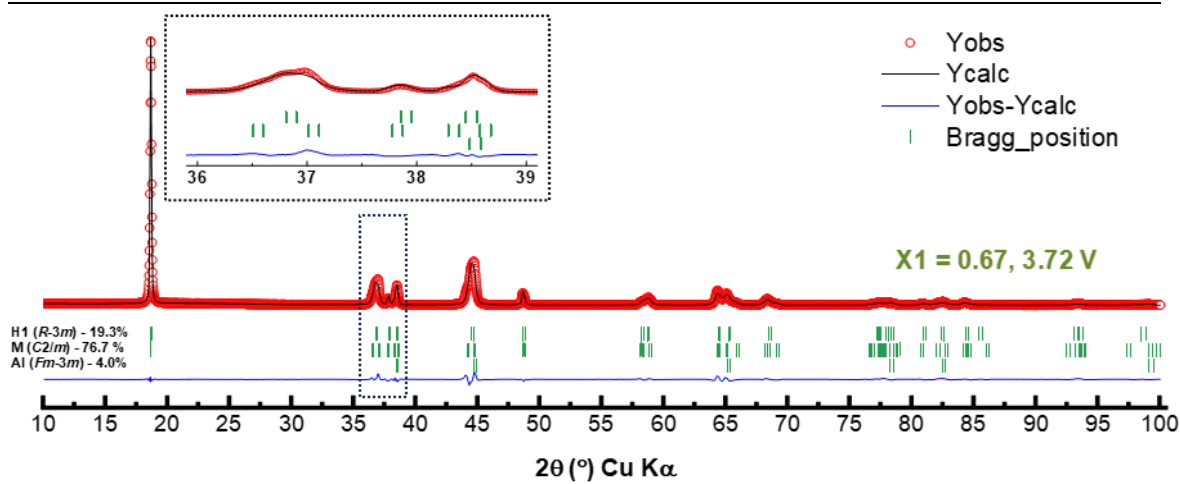

Table S1.2 Structural information including the lattice and atomic parameters of *ex situ* $Li_{0.50}NiO_2$ (X2) obtained from Rietveld refinement of the powder XRD data.

| X2 – $Li_{0.50}NiO_2$ | |
|---|---|
| Phase | Monoclinic – M |
| Space group | C 2/m |
| a = | 4.9610(4) Å |
| b = | 2.8312(3) Å |
| c = | 5.0685(4) Å |
| β | 109.47(6)° |
| V = | 67.12(9) Å³ |
| V/Z = | 33.56(2) Å³ |

| Atom | Wyckoff | x/a | y/b | z/c | Occ. | $B_{iso}$ (Å²) |
|---|---|---|---|---|---|---|
| Li | 2d | 0 | ½ | ½ | 0.503 | 0.95 |
| Ni1 | 2d | 0 | ½ | ½ | 0.033 | 0.95 |
| Ni2 | 2a | 0 | 0 | 0 | 1 | 0.53 |
| O1 | 4i | 0.7379(2) | 0 | 0.2134(6) | 1 | 0.86 |

| Distance | |
|---|---|
| Ni2–O | 1.9491(6) × 2 |
| Ni2–O | 1.9276(3) × 4 |
| Ni2–Ni2 | 2.8312(3) × 2 |
| Ni2–Ni2 | 2.8560(4) × 4 |
| Li(Ni1)-O | 2.1554(5) × 2 |
| Li(Ni1)-O | 2.1320(3) × 4 |

Agreement factors: $R_p$ = 6.13, $R_{wp}$ = 7.36, $\chi^2$ = 5.84

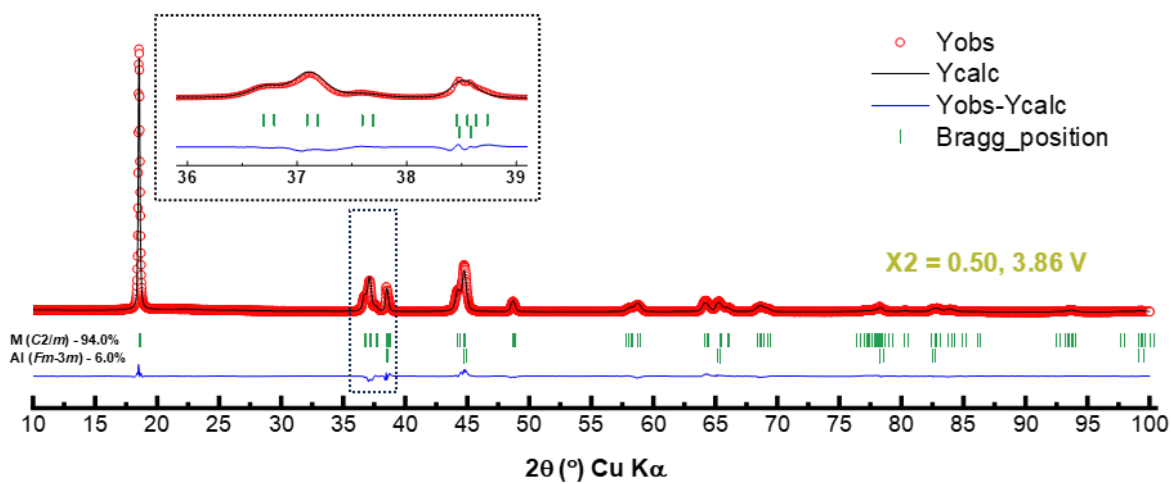

Table S1.3 Structural information including the lattice and atomic parameters of *ex situ* $Li_{0.34}NiO_2$ (X3) obtained from Rietveld refinement of the powder XRD data.

| X3 – $Li_{0.34}NiO_2$ | |
|---|---|
| Phase | Monoclinic – M |
| Space group | C 2/m |
| a = | 4.9310(4) Å |
| b = | 2.8178(2) Å |
| c = | 5.0918(5) Å |
| β | 109.17(6)° |
| V = | 66.83(2) Å³ |
| V/Z = | 33.42(2) Å³ |

| Atom | Wyckoff | x/a | y/b | z/c | Occ. | $B_{iso}$ (Å²) |
|---|---|---|---|---|---|---|
| Li | 2d | 0 | ½ | ½ | 0.343 | 0.95 |
| Ni1 | 2d | 0 | ½ | ½ | 0.033 | 0.95 |
| Ni2 | 2a | 0 | 0 | 0 | 1 | 0.53 |
| O1 | 4i | 0.7397(2) | 0 | 0.2092(6) | 2 | 0.86 |

| Distance | |
|---|---|
| Ni2–O | 1.9180(6) × 2 |
| Ni2–O | 1.9209(3) × 4 |
| Ni2–Ni2 | 2.8178(2) × 2 |
| Ni2–Ni2 | 2.8397(4) × 4 |
| Li(Ni1)-O | 2.1770(2) × 2 |
| Li(Ni1)-O | 2.1393(3) × 4 |

Agreement factors: $R_p$ = 6.98, $R_{wp}$ = 8.25, $\chi^2$ = 5.83

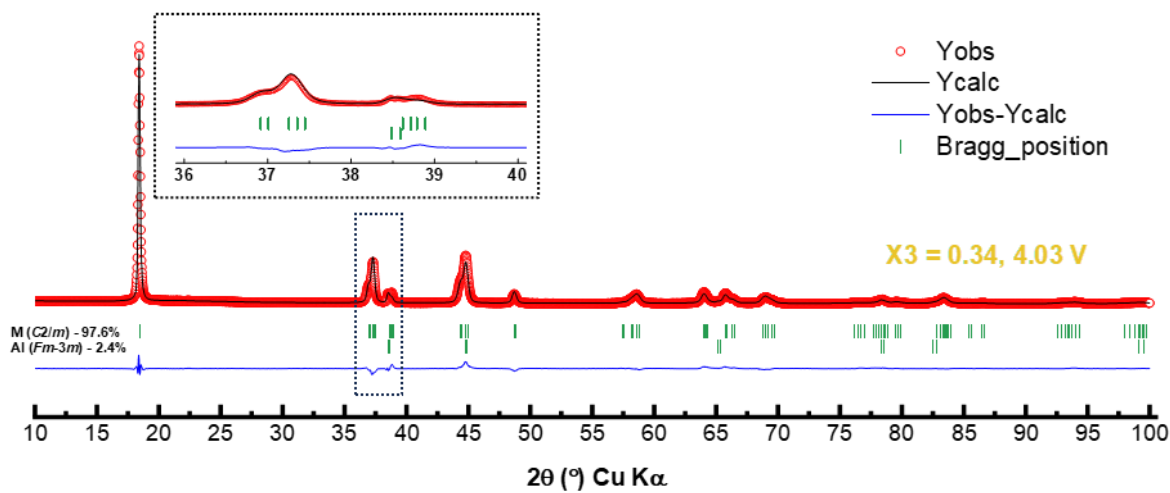

Table S1.4 Structural information including the lattice and atomic parameters of *ex situ* $Li_{0.27}NiO_2$ (X4) obtained from Rietveld refinement of the powder XRD data.

| X4 – $Li_{0.27}NiO_2$ | |
|---|---|
| Phase | Hexagonal – H2 |
| Space group | $R\bar{3}m$ |
| $a =$ | 2.8252(2) Å |
| $c =$ | 14.417(6) Å |
| $\alpha = \beta,\ \gamma$ | 90°, 120° |
| $V =$ | 99.66(2) Å$^3$ |
| $V/Z =$ | 33.22(2) Å$^3$ |

| Atom | Wyckoff | x/a | y/b | z/c | Occ. | $B_{iso}$ (Å$^2$) |
|---|---|---|---|---|---|---|
| Li | 3a | 0 | 0 | 0 | 0.270 | 0.78 |
| Ni1 | 3a | 0 | 0 | 0 | 0.033 | 0.78 |
| Ni2 | 3b | 0 | 0 | ½ | 1 | 0.15 |
| O1 | 6c | 0 | 0 | 0.2352(4) | 1 | 0.84 |

| Distance | |
|---|---|
| Ni2–O | 1.9068(5) × 6 |
| Ni2–Ni2 | 2.8252(2) × 6 |
| Li(Ni1)-O | 2.1595(8) × 6 |

Agreement factors: $R_p$ = 7.65, $R_{wp}$ = 8.19, $\chi^2$ = 7.21

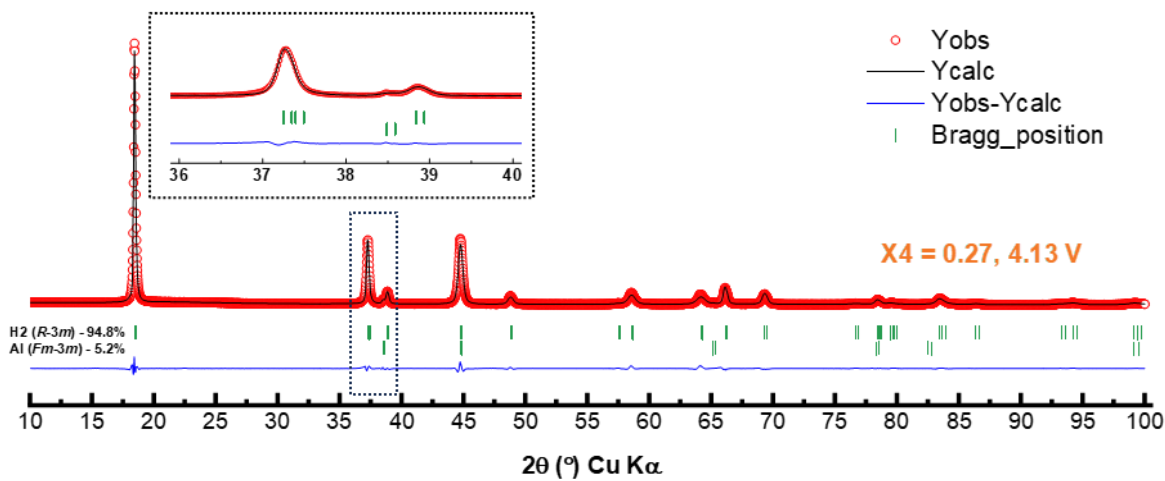

Table S1.5 Structural information including the lattice and atomic parameters of *ex situ* $Li_{0.03}NiO_2$ (X5) obtained from Rietveld refinement of the powder XRD data.

| X5 – $Li_{0.03}NiO_2$ | | | | | | |
|---|---|---|---|---|---|---|
| Phase | Hexagonal – H3, 99.6 wt. % | | | | | |
| Space group | $R\bar{3}m$ | | | | | |
| $a =$ | 2.8170(9) Å | | | | | |
| $c =$ | 13.467(2) Å | | | | | |
| $\alpha = \beta$, $\gamma$ | 90°, 120° | | | | | |
| $V =$ | 92.55(6) Å³ | | | | | |
| $V/Z =$ | 30.85(6) Å³ | | | | | |
| Atom | Wyckoff | x/a | y/b | z/c | Occ. | $B_{iso}$ (Å²) |
| Ni2 | 3b | 0 | 0 | ½ | 1 | 0.84(9) |
| O1 | 6c | 0 | 0 | 0.2381(9) | 2 | 0.32(2) |
| Bond | | | | Distance | | |
| Ni2–O | | | | 1.8897(3) x 6 | | |
| Ni2–Ni2 | | | | 2.8170(2) x 6 | | |
| Phase | Hexagonal – H2, 0.40 wt. % | | | | | |
| Space group | $R\bar{3}m$ | | | | | |
| $a =$ | 2.8252 Å | | | | | |
| $c =$ | 14.417 Å | | | | | |
| $\alpha = \beta$, $\gamma$ | 90°, 120° | | | | | |
| $V =$ | 99.66 Å³ | | | | | |
| $V/Z =$ | 33.22 Å³ | | | | | |
| Atom | Wyckoff | x/a | y/b | z/c | Occ. | $B_{iso}$ (Å²) |
| Li | 3a | 0 | 0 | 0 | 0.030 | 0.78 |
| Ni1 | 3a | 0 | 0 | 0 | 0.033 | 0.78 |
| Ni2 | 3b | 0 | 0 | ½ | 1 | 0.15 |
| O1 | 6c | 0 | 0 | 0.2352 | 1 | 0.84 |
| | | | Distance | | | |
| Ni2–O | | | | 1.9068 × 6 | | |
| Ni2–Ni2 | | | | 2.8252 × 6 | | |
| Li(Ni1)-O | | | | 2.1595 × 6 | | |

Agreement factors: $R_p$ = 14.5, $R_{wp}$ = 16.4, $\chi^2$ = 14.9

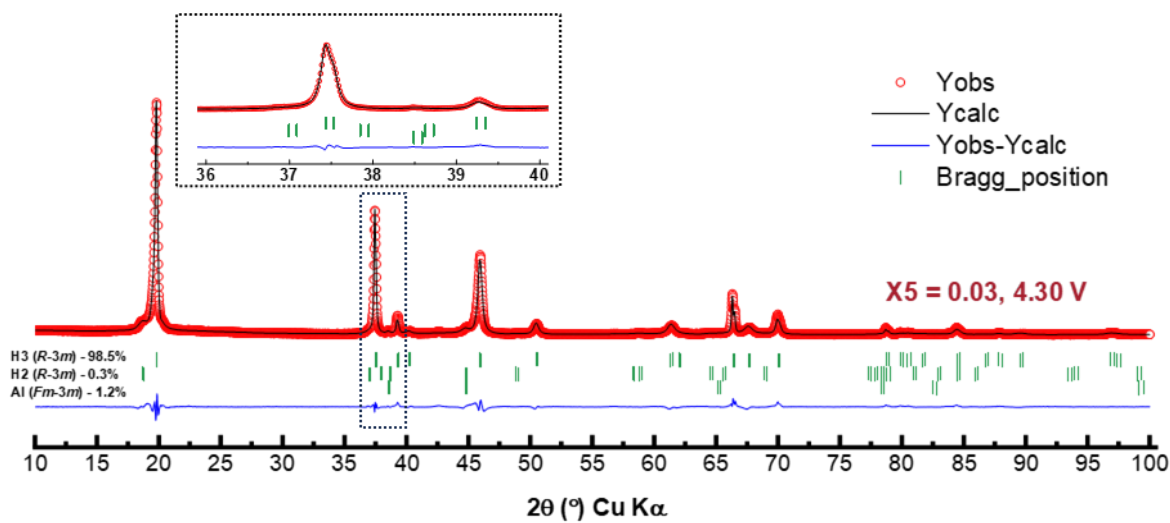

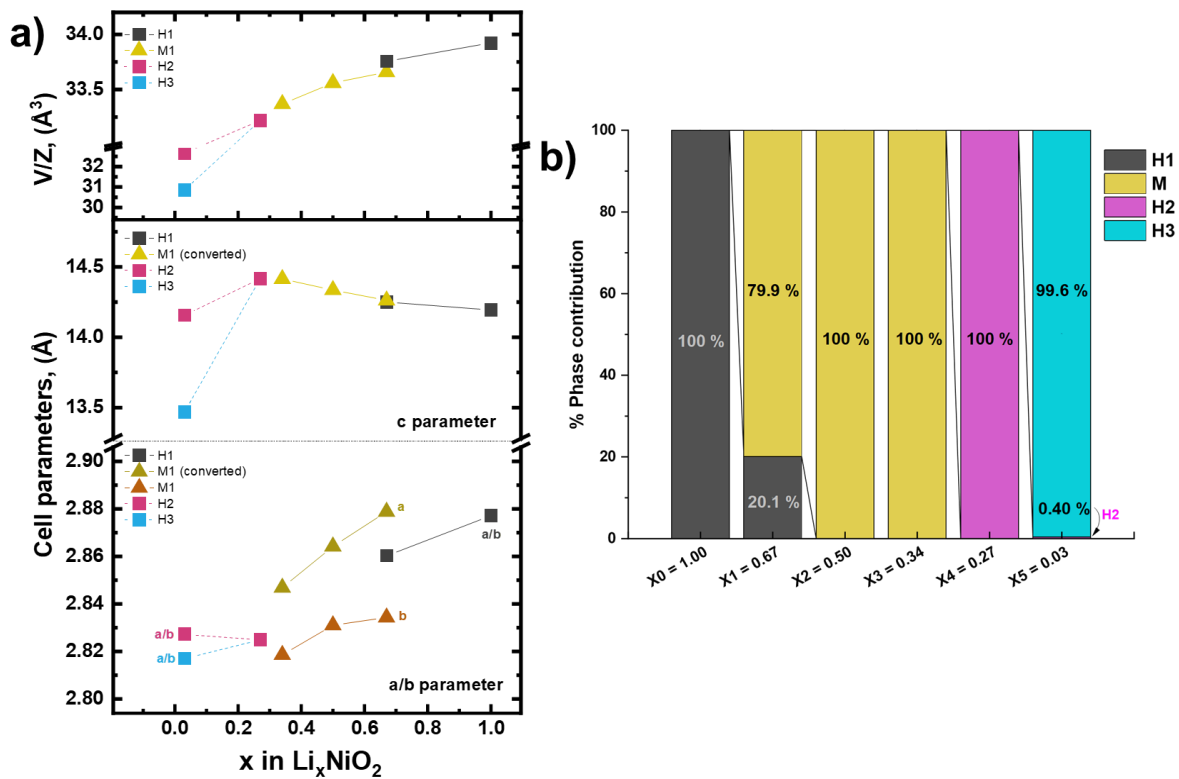

Figure S2. a) the volume and the cell parameters of the electrochemically prepared LNO. The values were obtained from the Rietveld refinements performed. Conversion of the c parameter in an hexagonal description in order to be able to compare the distance evolution perpendicular to the slabs stacking and using the formula $c_h = c_m * 3 * \sin\beta$, where $c_m$ is the c parameter in the monoclinic phase; b) summary of the phase contribution in each LNO sample. Reported values in main phase contribution are normalized without considering the contribution of Al.

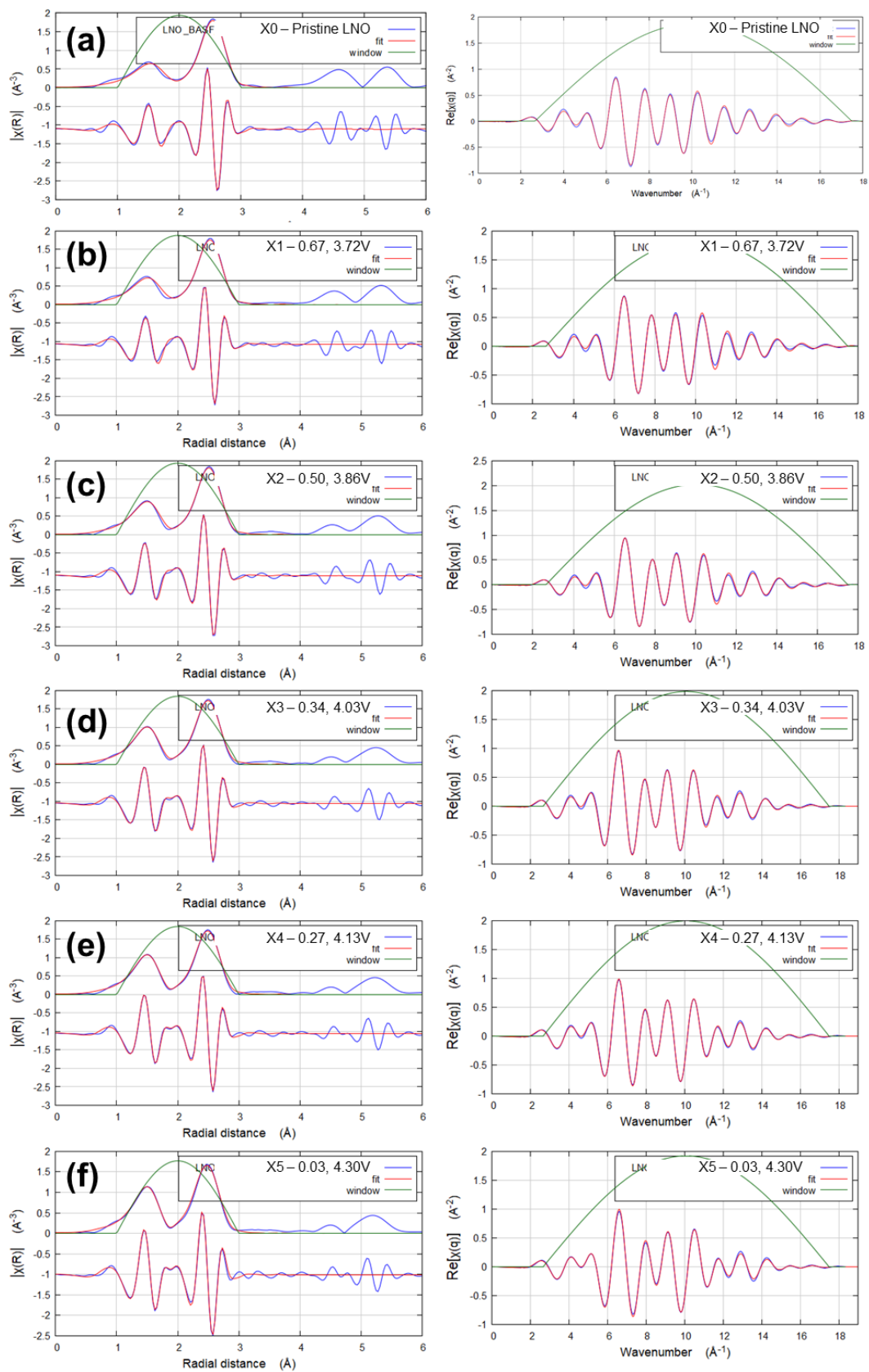

Figure S3. (left column) The magnitude of Fourier transforms of the EXAFS oscillations for pristine and *ex situ* deintercalated materials, fits of the Ni-O and Ni-Ni coordination shells are shown as red lines and compared to the experimental data as blue lines; and (right column) the *c*orresponding back Fourier transformed EXAFS oscillations in the q-space.

Table S2. Summary of the Ni—O, Ni—Ni, and Li—O distances obtained from the Rietveld refinements of the XRD data. Samples in hexagonal phase labelled, H, are distinguished from the monoclinic phase labelled, M, because of the 6 equidistant bonds.

| Sample - Formula | Average atomic distances determined from XRD | | | |
|---|---|---|---|---|
| | Crystal Phase | $d_{(Ni-O)}$, Å | $d_{(Ni-Ni)}$, Å | $d_{(Li-O)}$, Å |
| X0 - LiNiO$_2$ | H1 | 1.9771(5) × 6 | 2.8772(3) × 6 | 2.1060(5) × 6 |
| X1 - Li$_{0.67}$NiO$_2$ | H1 | 1.9480(6) × 6 | 2.8636(2) × 6 | 2.1320(9) × 6 |
| | M | 1.9610(9) × 2 | 2.8397(3) × 2 | 2.1480(9) × 2 |
| | | 1.9435(2) × 4 | 2.8691(6) × 4 | 2.1156(6) × 4 |
| X2 - Li$_{0.50}$NiO$_2$ | M | 1.9491(6) × 2 | 2.8312(3) × 2 | 2.1554(5) × 2 |
| | | 1.9276(3) × 4 | 2.8560(4) × 4 | 2.1320(3) × 4 |
| X3 - Li$_{0.34}$NiO$_2$ | M | 1.9180(6) × 2 | 2.8178(2) × 2 | 2.1770(2) × 2 |
| | | 1.9209(3) × 4 | 2.8387(4) × 4 | 2.1393(3) × 4 |
| X4 - Li$_{0.27}$NiO$_2$ | H2 | 1.9068(5) × 6 | 2.8252(2) × 6 | 2.1595(8) × 6 |
| X5 - Li$_{0.03}$NiO$_2$ | H2 | 1.9068(5) × 6 | 2.8252(2) × 6 | 2.1595(8) × 6 |
| | H3 | 1.8897(3) × 6 | 2.8170(2) × 6 | - |

Table S3. Summary of the average Ni—O and Ni—Ni distances obtained from EXAFS analysis.

| Sample - Formula | Crystal Phase from XRD | Average atomic distance from EXAFS | | | | |
|---|---|---|---|---|---|---|
| | | $d_{(Ni-O)}$, Å | MSRD, ($10^{-3}$ Å$^2$) | $d_{(Ni-Ni)}$, Å | MSRD, ($10^{-3}$ Å$^2$) | R-value, % |
| X0 - LiNiO$_2$ | Hexagonal | 1.95(1) | 8.72(1) | 2.88(1) | 4.18(1) | 0.70 |
| X1 - Li$_{0.67}$NiO$_2$ | Monoclinic | 1.88(1) × 4 | 4.76(1) | 2.87(1) | 4.38(1) | 0.23 |
| | | 2.00(1) × 2 | 7.04(1) | | | |
| X2 - Li$_{0.50}$NiO$_2$ | Monoclinic | 1.87(1) × 4 | 2.83(1) | 2.85(1) | 4.29(1) | 0.28 |
| | | 1.99(1) × 2 | 2.83(1) | | | |
| X3 - Li$_{0.34}$NiO$_2$ | Monoclinic | 1.87(1) × 4 | 2.81(1) | 2.84(1) | 4.60(1) | 0.31 |
| | | 1.96(1) × 2 | 2.81(1) | | | |
| X4 - Li$_{0.27}$NiO$_2$ | Hexagonal | 1.89(1) | 4.99(1) | 2.83(1) | 4.62(1) | 0.28 |
| X5 - Li$_{0.03}$NiO$_2$ | Hexagonal | 1.88(1) | 4.61(1) | 2.82(1) | 4.87(1) | 0.60 |

*MSDR – Mean square relative displacement

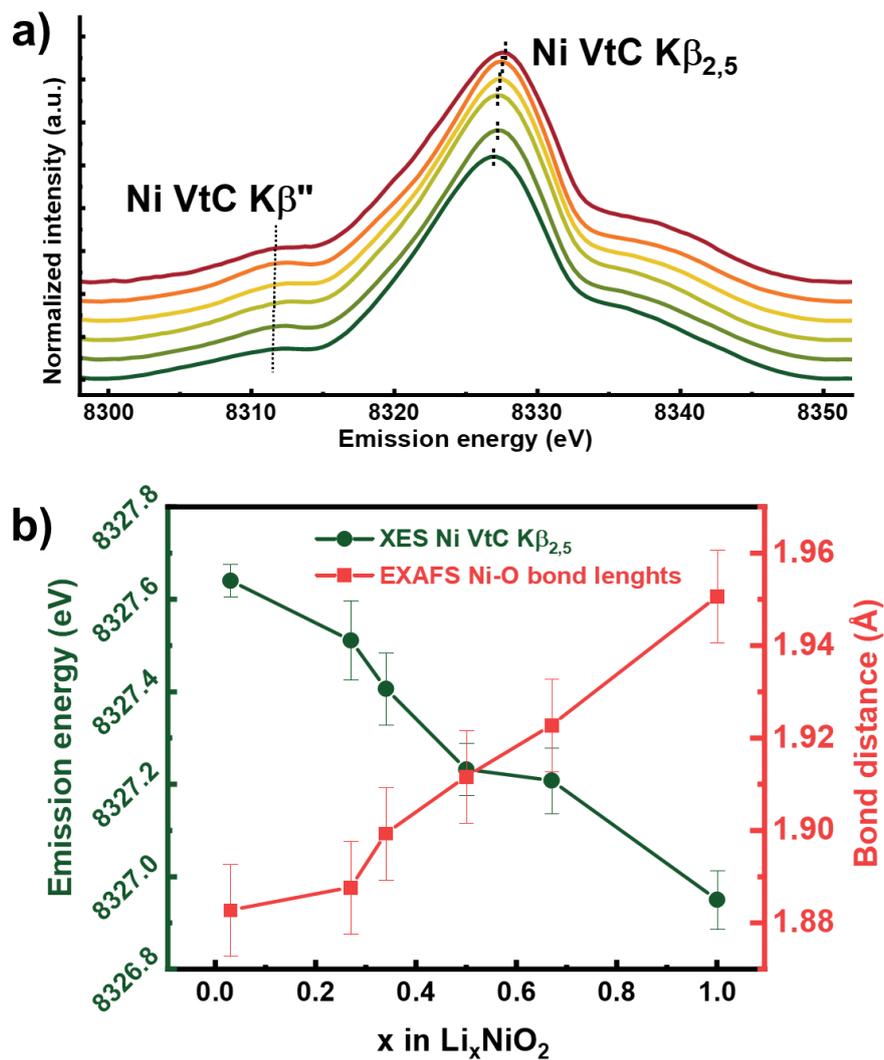

Figure S4. a) The VtC Kβ XES spectra and b) the anticorrelation between the main peak position (~8325 eV) of VtC Kβ emission line and the Ni—O bond distance as a function of Li$^+$ content.